\documentclass[12pt]{article}
\usepackage{soul}
\usepackage{xcolor}

\usepackage{scicite}

\usepackage{times}

\usepackage{physics}
\usepackage{lineno}

\usepackage{graphicx}
\usepackage{capt-of}%

\usepackage{amsmath}
\usepackage{amssymb}

\usepackage{float}
\usepackage{chngcntr}
\usepackage{tikz}

\usepackage{epsf}

\newenvironment{sciabstract}{%
\begin{quote} \bf}
{\end{quote}}

\title{Electric field tunable unconventional superconductivity in  alternating twist magic-angle trilayer graphene}
\author
{Zeyu Hao$^{1\dagger}$, A. M. Zimmerman$^{1\dagger}$, Patrick Ledwith$^{1}$, Eslam Khalaf$^{1}$,\\Danial Haie Najafabadi$^{1}$, Kenji Watanabe$^2$, Takashi Taniguchi$^3$,\\ Ashvin Vishwanath$^{1}$ \& Philip Kim$^{1*}$\\\\

\normalsize{$^1$ Department of Physics, Harvard University,}\\ 
\normalsize{Cambridge, Massachusetts 02138, USA}\\
\normalsize{$^2$ Research Center for Functional Materials, National Institute for Material Science,}\\
\normalsize{1-1 Namiki, Tsukuba 305-0044, Japan}\\
\normalsize{$^3$ International Center for Materials Nanoarchitectonics, National Institute for Material Science,}\\
\normalsize{1-1 Namiki, Tsukuba 305-0044 Japan}\\
\normalsize{$\dagger$Theses authors contribute equally to this work.}\\\\
\normalsize{$^*$Correspondence and requests for materials should be addressed to:}\\
\normalsize{P.K.~(email: pkim@physics.harvard.edu).}}
\date{}
\begin{document}
\baselineskip24pt
\maketitle
\pagebreak
\begin{sciabstract}

We construct a van der Waals heterostructure consisting of three graphene layers stacked with alternating twisting angles $\pm\theta$. At the average twist angle $\theta\sim 1.56^{\circ}$, a theoretically predicted magic angle for the formation of flat electron bands, narrow conduction and valence moir\'e bands together with a linearly dispersing Dirac band appear. Upon doping the half-filled moir\'e valence band with holes, or the half-filled moir\'e conduction band with electrons, displacement field tunable superconductivity emerges, reaching a maximum critical temperature of 2.1~K. By tuning the doping level and displacement field, we find that superconducting regimes occur in conjunction with flavor polarization of moir\'e bands bounded by a van Hove singularity at high displacement fields. These experimental results are inconsistent with a weak coupling description, suggesting that the observed moir\'e superconductivity has an unconventional nature.  
 
\end{sciabstract}

The experimental realization of twisted bilayer graphene (TBG) opened up new possibilities for studying interaction effects in moir\'e engineered electronic bands. It was first predicted theoretically that the hybridization of two twisted graphene sheets could produce nearly flat bands at the so called ``magic angles" (MA) \cite{Bis.11,San.07,Sha.10,Sua.10}. Initial experiments showed that a significant reduction of kinetic energy gives rise to correlated insulating phases and superconductivity upon doping these insulating states\cite{cao_correlated_2018,Cao.18}. In the followup experiments, additional interaction-driven phases were discovered in MA-TBG, including isospin symmetry breaking metals\cite{Sai.20b}, orbital ferromagnetism\cite{Sha.19,Lu.19,Ser.20}, and magnetic field induced Chern insulators\cite{Wu.20,nuckolls_strongly_2020,Sai.20}. Despite the rapid progress of the field, the question of whether the superconductivity is unconventional, driven by strong electron-electron interactions, 
or conventional, arising from electron-phonon interaction at weak coupling,  
remains under debate. Some experiments suggested that the superconducting and insulating phases are independent or maybe even competing, with the superconductivity persisting or strengthening when Coulomb interaction is screened \cite{Ste.20,Sai.20c,Aro.20,Liub.20}. However, others have provided evidence that the superconductivity has unconventional features, such as coexisting nematic order and a lack of correlation between large density of states and higher critical temperature~\cite{park2020flavour}, pointing to unconventional superconductivity of non-phononic origin\cite{Cao.20}. 

The creation of moir\'e engineered van der Waals (vdW) interfaces has been extended to other 2-dimensional (2D) material systems as well, leading to the observation of many interesting interaction-driven phases, such as quantum anomalous Hall states in twisted monolayer- bilayer graphene\cite{Pol.20,Che.20} and generalized Wigner crystal in WSe$_2$/WS$_2$ moir\'e superlattices\cite{Reg.20,xu_correlated_2020}. On the other hand, MA-TBG remains the only system where superconductivity is unambiguously well-established  \cite{Cao.18,Yan.19,Lu.19,Aro.20}. In contrast, initial reports of superconductivity in other 2D flat band systems such as ABC trilayer graphene aligned with BN\cite{Che.19}, twisted double bilayer graphene\cite{Liu.20,he_symmetry_2020}, 
and twisted WSe$_2$\cite{An.20} have proven less conclusive\cite{Bal.20}.

In this work, we study a new type of moir\'e engineered graphene multi-layer system, MA twisted trilayer graphene (TTG)  with vertical mirror symmetry\cite{Kha.19}. We present a clear signature of superconductivity controlled by applied electric field. The continuously tunable band structure of MA-TTG provides a new experimental knob for probing the superconducting mechanism. 

Our TTG consists of three layers of graphene, with the twist angle between the top layer (T) and middle layer (M) being $\theta$, and the twist angle between the middle layer and the bottom layer (B) being $-\theta$ as shown in the schematic in Fig.~1a. The stacking with this alternating sequence of angles with opposite signs preserves the vertical mirror plane symmetry (see Supplementary Material (SM), S1), differing from the previously studied trilayer systems \cite{Tsa.20}.  A recent theoretical work predicted that the Hamiltonian for this system can be effectively decoupled into that of a monolayer graphene and a TBG with the inter-layer coupling strength enhanced by a factor of $\sqrt{2}$\cite{Kha.19}. As a result, the band structure of TTG consists of a Dirac cone from the monolayer graphene coexisting with TBG flat bands. Interestingly, the magic angle is predicted to be the TBG magic angles multiplied by $\sqrt{2}$: $\theta_{TTG}=1.56^\circ$ \cite{Kha.19}. Fig.~1c shows a band structure of TTG at this angle. To experimentally realize such a system, we utilize the ``cut and twist" technique\cite{Sai.20c} (see SM S1 for detail). An image of the completed device is shown in Fig.~1b. The colored lines trace the original positions of the three pieces of graphene. In addition to the TTG device fabricated in the three-layer overlapped region (TMB), we made two other TBG devices,  one in the region with only the top  and middle layers (TM), and the other in the region where there are only the middle and bottom layers (MB).  These two devices allow us to measure the TM twist angle, $\theta_{\text{TM}}$, and MB twist angle, $\theta_{\text{MB}}$, individually to characterize our devices.

Fig.~1d and e show the longitudinal resistivity, $\rho$, as a function of perpendicular magnetic field ($B$) and carrier density $n$, controlled by both top and bottom gates of the two TBG devices, TM and MB. They exhibit typical magnetotransport features of large angle TBG\cite{Cao.16}, with no insulating resistivity peaks other than at $\nu=0$ and 4. Here, $\nu$ is the moir\'e band filling factor $\nu=4n/n_s$, where $n_s$ is the carrier density at full filling of the four-fold degenerate moir\'e bands. From these fan diagrams, we estimate  $\theta_{\text{TM}}=1.35^\circ$ and $\theta_{\text{MB}}=-1.69^\circ$. The difference in angles is expected due to imperfect angle control in experiments as well as the ubiquitous angle disorder in twisted devices. Interestingly, however, we find such small angle difference between TM and MB regions yields no appreciable effect in the TTG device formed in the TMB region. Fig.~1f shows the Landau fan diagram of the TTG formed in the TMB region at a fixed back gate voltage, $V_{\text{BG}}=0$. $\rho(B,n)$ exhibits emergent Landau fans at integer fillings $\nu=0, \pm2, 1, 3$ which correspond to a twist angle of 1.55$^\circ$. In addition, the device is highly uniform across most pairs of contacts, with angle disorder on the order of $0.02^\circ$ estimated from similar magneto-transport data (SM, S3). Such high uniformity of twist angle in the TMB region might indicate that the strain relaxation on atomic length scales forces $\theta_{\text{TM}}=-\theta_{\text{MB}}$ to be the average twist angle, favoring alignment between the top and bottom layer to reduce the structural energy\cite{Car.20}. An alternative scenario is that the three layers are coupled so strongly that they behave as a single system with the measured angle the average of $\theta_{\text{TM}}$ and $\theta_{\text{MB}}$. 
In either case, the resulting uniform device indicates that TTG is relatively robust against small angle misalignment and disorder. 

We find that, unlike in MA-TBG samples, $\rho(n)$ measured in the TMB device at $B=0$ (Fig. 1g) does not exhibit strong insulating behavior at any filling, consistent with the additional Dirac cone in the proposed TTG band structure in Fig. 1c. The Hall conductivity $\sigma_{xy}$ data obtained from the TTG device also confirms the proposed single particle band structure of TTG. The Landau fan emanating from the charge neutrality has sequence $-2, -6, -10, -14$ on the hole doped side and 2, 4, 6 on the electron doped side. This sequence indicates that Landau levels are four-fold degenerate but is different from the typical 4, 8, 12, 16 sequence obtained in MA-TBG\cite{Cao.18}.  This change is similar to the sequence in ABA trilayer graphene \cite{Hen.12}, and likely results from the presence of the Dirac cone. Flavor symmetry breaking is evident in the  Landau fan coming from $\nu=-2$ with sequence $-2, -4, -6, -8$ showing only two-fold degeneracy. The resistive states at  $\nu=\pm2$ show slopes consistent with Chern number $C=\pm2$. Fig.~1h shows $\sigma_{xy}(n)$ taken at $B=10$~T. Interestingly, we see large regions of $C=-2$ near $\nu=-4$ and $C=2$ near $\nu=4$ (also see $\sigma_{xy}(B,n)$ in SM, S4). More direct experimental evidence is present in the Landau fan diagram taken at zero-displacement field, where we observe quantum Hall sequences directly originating from the Dirac cone (see SM S4).


At low temperature and magnetic field, we find large regions of robust superconductivity in the TTG sample. Fig.~2a shows $\rho(n)$ at different temperatures at $V_{\text{BG}}=-4.5$~V. At our lowest experimental temperature of $T = 0.34$~K, zero resistance regions appear on the hole-doped side of $\nu=-2$ and electron-doped side of $\nu=2$. The two insets show $\rho(n,T)$  near $\nu=-2$ and $\nu=2$ respectively, both displaying clear superconducting domes. The transition in $\rho(T)$ across the dome boundary is sharp as shown in Fig.~2b at  $\nu=-2.3$, the optimal filling for $\nu<-2$. It is noteworthy that $\rho=0$ at $T\sim2.1$~K, which is higher than most MA-TBG devices in published literature\cite{Cao.18,Yan.19,Lu.19,Aro.20}. At $\nu=-2.3$, we measure a Berezinskii–Kosterlitz–Thouless (BKT) transition of $2.16$~K from  the power law dependence of the current and voltage $I$-$V$ characteristics as shown in the inset of Fig.~2b. Phenomenologically, we characterize, $T_c$, as the temperature at which $\rho$ falls to $10\% $ of the normal state resistance, $\rho_N$, which we find to be consistent with the BKT transition temperature and a better measure for 2D superconductivity (SM, S5). 
Additional clear signatures of superconductivity are also visible in the differential resistance, $dV/dI$, as a function of DC bias current as shown in Fig.~2c, which shows a sharply defined critical current $I_c$. At $B=0$~T, the sudden increase of $dV/dI$ occurs at $I_c=\pm720$~nA. As the magnetic field increases, $I_c$ becomes smaller and the shape of $dV/dI$ becomes more smooth, a characteristic behavior of 2D superconducivity  suppressed by  perpendicular magnetic field. The resulting critical field, $B_c$, is evident in $\rho(T,B)$ in Fig.~2d. We extract the Ginzburg-Landau (GL) coherence length $\xi_{\text{GL}}$ from the  theory for a 2D superconductor: $B_c=[\Phi_0/(2\pi\xi^2_{\text{GL}})](1-T/T_c)$ \cite{Tin.04}, where $\Phi_0$ is the superconducting flux quantum. Using the BKT transition temperature as $T_c$ in the above relation, we estimate $\xi_{\text{GL}}=61$~nm,  several times the interparticle distance estimated from $n$. Using $T_c$ extracted at $\rho=0.5\rho_N$, considering prevailing fluctuation effects, we find $\xi_{\text{GL}}=13.4$~nm, only slightly larger than the interparticle distance.

Employing both the top and bottom gates, we can control both $n$ and the displacement field $D$ independently, tuning the superconductivity in TTG by the electric field. Fig.~3a shows $\rho$ as a function of $\nu$ and displacement field $D$ at temperature $T=0.34$~K. We observe at charge neutrality a resistive peak that is not disturbed by $D$. This is expected for the Dirac cone crossings in the flat bands. At $\nu=\pm$2, 1 and 3, there are resistivity peaks that are modulated by $D$. At $\nu=\pm$4, the system has low $\rho$, which is expected due to the existence of the additional Dirac cone and lack of band insulators at full filling. The superconductivity appears as the dark blue regions both on the hole side between $\nu=-3$ and $\nu=-2$, and on the electron side between $\nu=2$ and $\nu=3$. The hole side superconductivity persists for all $D$, with a width that first increases with $D$, and starts to decrease at $D/\epsilon_0\sim0.4$~V/m at our base temperature. The electron side superconductivity is weaker and affected more strongly by $D$. At $T=0.34$~K, it only starts to emerge at  $D/\epsilon_0\sim-0.4$~V/nm. To better illustrate the evolution of the superconductors with $D$, we have measured $\rho(n, T)$ at several discrete $D$s, as shown in Fig.~3b--d for holes and Fig.~3e--g for electrons, showing dome-like superconducting regimes (several representative $\rho(T)$ curves are shown in Fig.~3a insets). While the optimal dopings where the maximum $T_c$ occurs, $\nu_{op}\approx \pm 2.4$, is insensitive to $D$, we find the maximum $T_c$ of the dome and the filling range, $\Delta \nu$, i.e., the height and width of the dome, are sensitive to $D$. We measure $\rho(T)$ at different $D$  at optimal filling to extract the transition temperature $T_c$ at each $D$, providing a quantitative description of the $D$ dependence of superconductivity. Fig.~3h and i show transition temperature as a function of $D$ for the hole side and electron side superconductors respectively.  For the hole side superconductor, starting from $D/\epsilon_0=0$, $T_c$ first increases, reaches maximum at around $D/\epsilon_0=0.4$~V/nm and then decreases quickly. The electron side superconductor displays a similar trend, with $T_c$ increasing after appearing at $ D/\epsilon_0\sim-0.5$~V/nm then decreasing  below $D/\epsilon_0=-0.62$~V/nm.

The electric field tunable superconductivity in TTG can be ascribed to the tuning of single particle bands controlled by $D$. Fig. 4a shows Hall density ($n_H=\sigma_{xy}B/e$, $e$ is the electron charge) at a low magnetic field $B=0.5$~T, near the region where the hole side superconductor resides. At  $D/\epsilon_0=0.2$~V/nm, away from zero filling, $n_H(\nu)$ increases linearly with a unity slope and then resets to 0 near $\nu=-2$. This reseting behavior has been considered as a signature of the spin and valley isospin symmetry breaking, where the four-fold degeneracy turns into two-fold. After this flavor symmetry breaking, the electrons completely fill the two lower energy bands and $n_H$ corresponds to the density in the two higher energy bands. Similar flavor symmetry breaking in moir\'e flat bands has been observed and discussed in MA-TBG\cite{Sai.20c,Wu.20,Won.20,Zon.20}. This symmetry breaking can be better illustrated by the quantity $n_H-\nu$, which directly gives the degeneracy of the symmetry breaking phase\cite{Sai.20b}. Fig.~4d shows $|n_H-\nu|$ as a function of $\nu$ and $D$, showing several symmetry breaking regions. Above $\nu=-2$, the large area of $|n_H-\nu|=0$ shown as dark blue indicates that the holes are filling the four bands equally. Between $\nu=-2$ and $\nu=-3$, the system enters the symmetry breaking phase with two degenerate bands where $|n_H-\nu|=2$, shown in white. At $\nu=-3$ at small $D$ another reset occurs, below which $|n_H-\nu|$ is not integer valued,  changing gradually from 3 to 4. 

As $D$ tunes the single particle band structure of the MA-TTG, the flavor symmetry breaking also changes. For example, for the hole side band ($\nu<0$), above $D/\epsilon_0=0.35$~V/nm a large region $|n_H-\nu|=4$ emerges below $\nu<-3$ (marked as I in Fig. 4~b), indicating that the four bands are being filled equally with no symmetry breaking. Interestingly, we find this region is bounded on the right by a van Hove singularity (vHS), whose existence can be detected from diverging $n_H$ followed by a sign change\cite{Wu.20}. The characteristic sharp divergences of two vHSs can be seen in Fig.~4a at $D/\epsilon_0=0.4$~V/nm near $\nu=-3$ (marked by vertical arrows), combining into one large divergence at larger $D$. The left boundary of region I also shows a discontinuity. However, $n_H$ value across this boundary is continuous, indicating that this is not a vHS. As $D$ increases, the flavor-polarizing vHS moves to the right, expanding the $|n_H-\nu|=4$ region. Importantly we note that this evolution correlates with the reduction of superconductivity. When the boundaries of the zero magnetic field superconducting region is superimposed onto the $|n_H-\nu|$ plot, shown in Fig.~4b (dashed lines), we can see that the superconducting region is reduced as the vHS and $|n_H-\nu|=4$ region crowd out the $|n_H-\nu|=2$ region. For the electron side superconductor, where superconductivity only is visible at finite $D$, similar analysis (Fig. 4d) shows that the superconducting region also shrinks when the vHS starts to cross the symmetry breaking phase boundary.

The region near a vHS has an increased  density of states (DOS), which promotes superconductivity in conventional Bardeen-Cooper-Schrieffer (BCS) theory in the weak coupling limit\cite{Bar.57,Och.18}. Here, instead it is observed that superconductivity weakens as a vHS approaches and subsequently flavor polarization occurs.
The prominent role of vHS in the system can also be captured in single particle band calculations. Fig.~4g shows calculated DOS as a function of filling and interlayer electric potential $U$, which is directly proportional to the experimental $D$. The calculated DOS is symmetric between positive and negative $U$ so only the positive part is shown (see SM S2 for details). We observe that at low $D$, there is high density of states concentrated near charge neutrality, which is a  reflection of the flatness of the bands.  An example band structure at low $D$ with $U=11$~meV is shown in Fig.~4e. As $D$ increases, the bands become more dispersive and vHSs become prominent, shown as the white lines in the DOS calculation at larger $U$. An example band structure in this range is shown in Fig.~4f. The prominent vHS in the theoretical density of states at large $U$ agrees with the vHS that appear at large $D$ in experiments. 

The intrusion of this vHS, and the subsequent flavor ordering which limits the width of the $|n_H-\nu|=2$ region and consequently the region of superconductivity, accounts for the reduction of the superconducting dome at large $D$. However,  the initial enhancement of the superconductivity seems to lie in the region where band flatness dominates the physics.
In this regime, the average DOS and bandwidth of the individual conduction and valence band remains roughly constant as shown in Fig.~4g and h. The major change in the single particle band structure in this small $D$ range happens at the $K$ point, where the conduction and valence bands gradually split away from each other, increasing the combined bandwidth at this point. Recent theoretical work has suggested the importance of a second order process coupling flat bands, reminiscent of super-exchange, as the driving force for pairing\cite{Bul.20,Kha.20}.
This process leads to an energy scale $J\sim t^2/E_c$, where $t$ is related to the overall effective bandwidth, while $E_c$ is a measure of the repulsion. 
This pairing mechanism also invokes the presence of $C_{2z}T$ symmetry. Indeed, this symmetry requirement is consistent with the fact that, at present, MA-TBG and alternating MA-TTG are the only two platforms exhibiting robust superconductivity, and they are also unique among existing moir\'e systems in retaining this symmetry. 
Within this picture, changing the overall effective bandwidth $t$ can enhance superconductivity, which may be related to the observed  enhancement of both bandwidth and $T_c$ on increasing the displacement field at small $D$. 
Further  evidence for the strong coupling nature of superconductivity is provided by the rapid increase of $T_c(\nu)$ with doping observed in the superconducting domes for $|\nu|<|\nu_{op}|$. This suggests a picture where tightly bound Cooper pairs condense leading to $T_c$ which is limited by density and therefore grows with doping.  One possible strong coupling mechanism that is broadly consistent with these observations is  skyrmion superconductivity\cite{Kha.20}, wherein the $J$ interaction binds charged skyrmions into pairs. We estimate $J$ to be a few meV from the experimentally obtained slope of $T_c(\nu)$ in the regime of $|\nu|<|\nu_{op}|$, consistent with theoretical expectations\cite{Kha.20} (SM,S6). We anticipate these results to stimulate further theoretical and experimental investigations into these novel correlation driven phenomena.  

\section*{Acknowledgments}
We thank Pablo Jarillo-Herrero, Xiaomeng Liu, Yuval Ronen and Onder Gul for fruitful discussions. The major experimental work is supported by NSF (DMR-1922172). P.K. acknowledges support from the DoD Vannevar Bush Faculty Fellowship N00014-18-1-2877. Z.H. is supported by ARO MURI (W911NF-14-1-0247).  AV and EK were supported by a Simons Investigator award (AV)  and by the Simons Collaboration on Ultra-Quantum Matter, which is a grant from
the Simons Foundation (651440, A.V.). PL was supported by the Department of Defense (DoD) through the National Defense Science \& Engineering Graduate Fellowship (NDSEG) Program.
K.W. and T.T. acknowledge support from the Elemental Strategy Initiative conducted by the MEXT, Japan, Grant Number JPMXP0112101001, JSPS KAKENHI Grant Number JP20H00354 and the CREST(JPMJCR15F3), JST.

\bibliography{TTG.bib}
\bibliographystyle{naturemag_noURL}

\clearpage
\begin{figure}
\centering
\includegraphics[width=\textwidth]{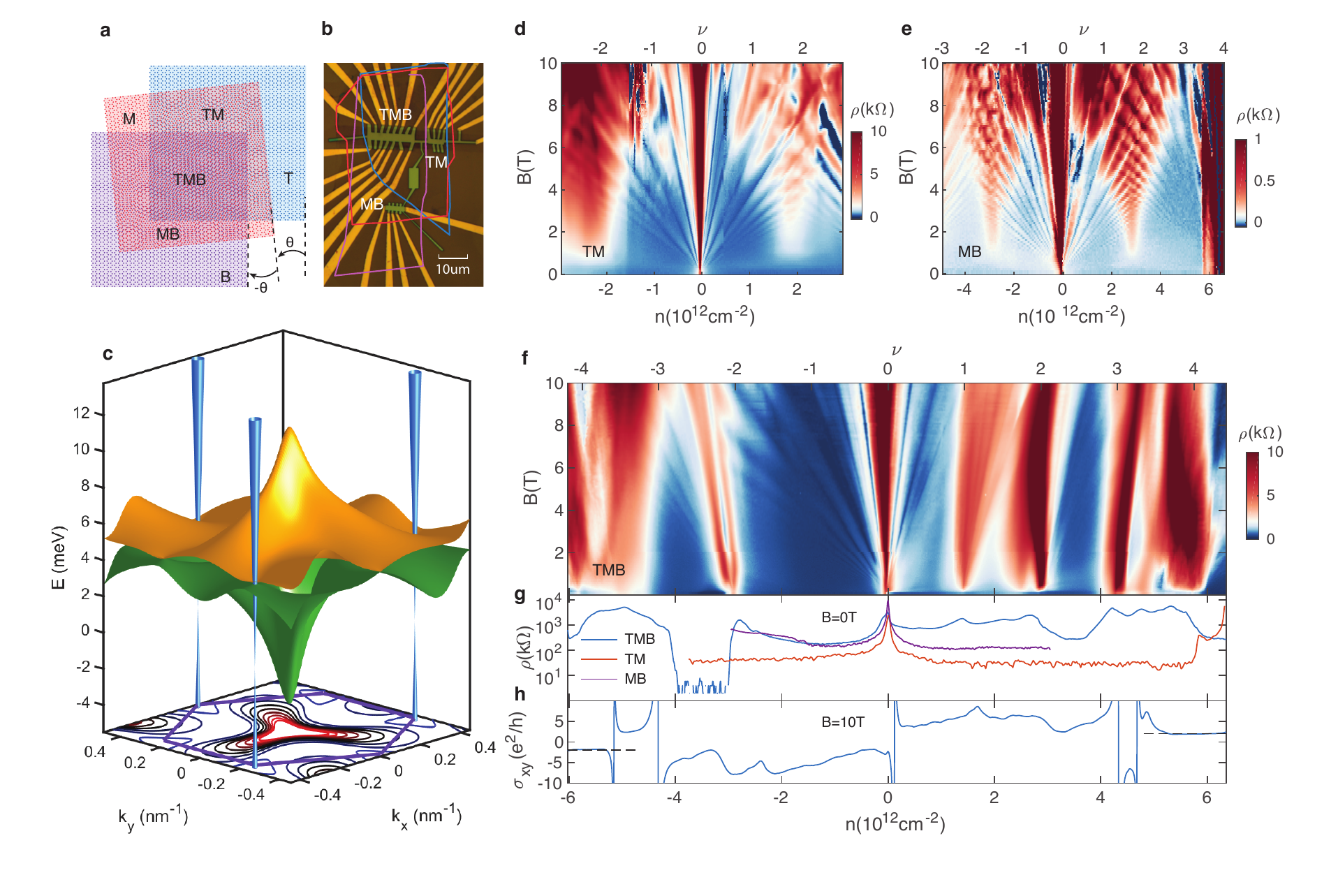}
\caption{ \textbf{$\vert$ Device structure and characterization} \textbf{a,} Schematic diagram of the three layers of TTG with an alternating twist angle $\theta$. The top (T) and bottom (B) layers are aligned while the middle layer (M) it twisted by $\theta$ relative to both layers, preserving the in plane mirror symmetry.  \textbf{b,} Optical microscope image of the TTG device fabricated in the TMB region and two TBG devices fabricated in the TM and MB regions.\textbf{c} Theoretical band structure for MA-TTG at $D/\epsilon_0=0$ plotted on the mini Brillouin zone (BZ) marked in purple in the bottom face. The blue Dirac cones sit at the mini BZ $K$ points, while the flat bands, orange (conduction) and green (valence) are the most dispersive at the mini BZ $\Gamma$ point. A contour plot of the valence band is projected on the $x$--$y$ plane. \textbf{d,e} Landau fan diagrams of the two TBG hall bars TM (\textbf{d}) and MB (\textbf{e}) In each device fans are visible emanating from $\nu=0$ and $\nu=\pm4$ as well as an increase in resistance at the vHS near $\nu=\pm2$. \textbf{f,} 
Landau fan diagram of the TTG hall bar (TMB). Resistive states and fans emerge at $\nu=0$, $+1$, $\pm2$, $+3$, and near $\pm4$. \textbf{g,} Zero magnetic field resistivity as a function of filling in TM, MB and TMB. \textbf{h,} Hall conductivity in TMB at $B=$10~T.}
\end{figure}

\begin{figure}
\centering
\includegraphics[width=\textwidth]{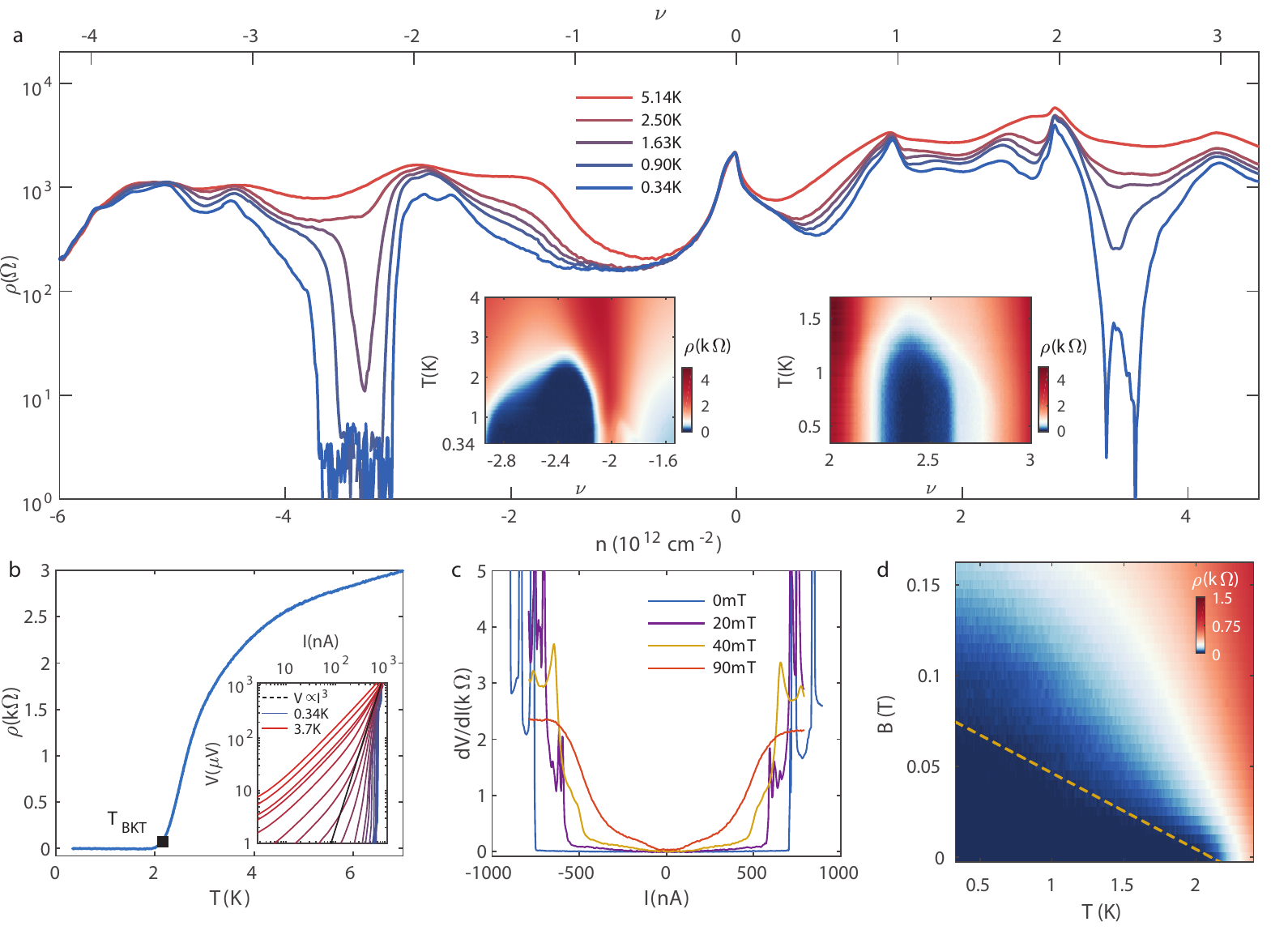}
\caption{ \textbf{$\vert$ Superconductivity in TTG} \textbf{a,} $\rho$ as a function of $\nu$ taken at a fixed $V_{BG}=-4.5$~V at several different temperature values. The formation of superconducting regions is visible at $\nu<-2$ and $\nu>2$. The left (right) inset shows the superconducting domes in the the $T-\nu$ plane at $D/\epsilon_0=-0.55$~V/nm for $\nu>2$ at taken along a cut along $V_{BG}=0$~V for $\nu<-2$. \textbf{b,} Superconducting transition in  resitivity at $\nu=-2.3$ and $D/\epsilon_0=0.29$~V/nm. The BKT transition temperature is marked where $V\propto I^3$ as shown in the inset.  \textbf{c,} Differential resistance measurements as a function of bias current at different magnetic fields.  \textbf{d,} $\rho$ as a function of temperature and magnetic field at $\nu=-2.3$ and $D/\epsilon_0=0.4$ V/nm. The dashed line corresponds to a GL theory fit with a coherence length of $\xi_{\text{GL}}=61$ nm (see SM for more detail).
}
\end{figure}

\begin{figure}
\centering
\includegraphics[width=\textwidth]{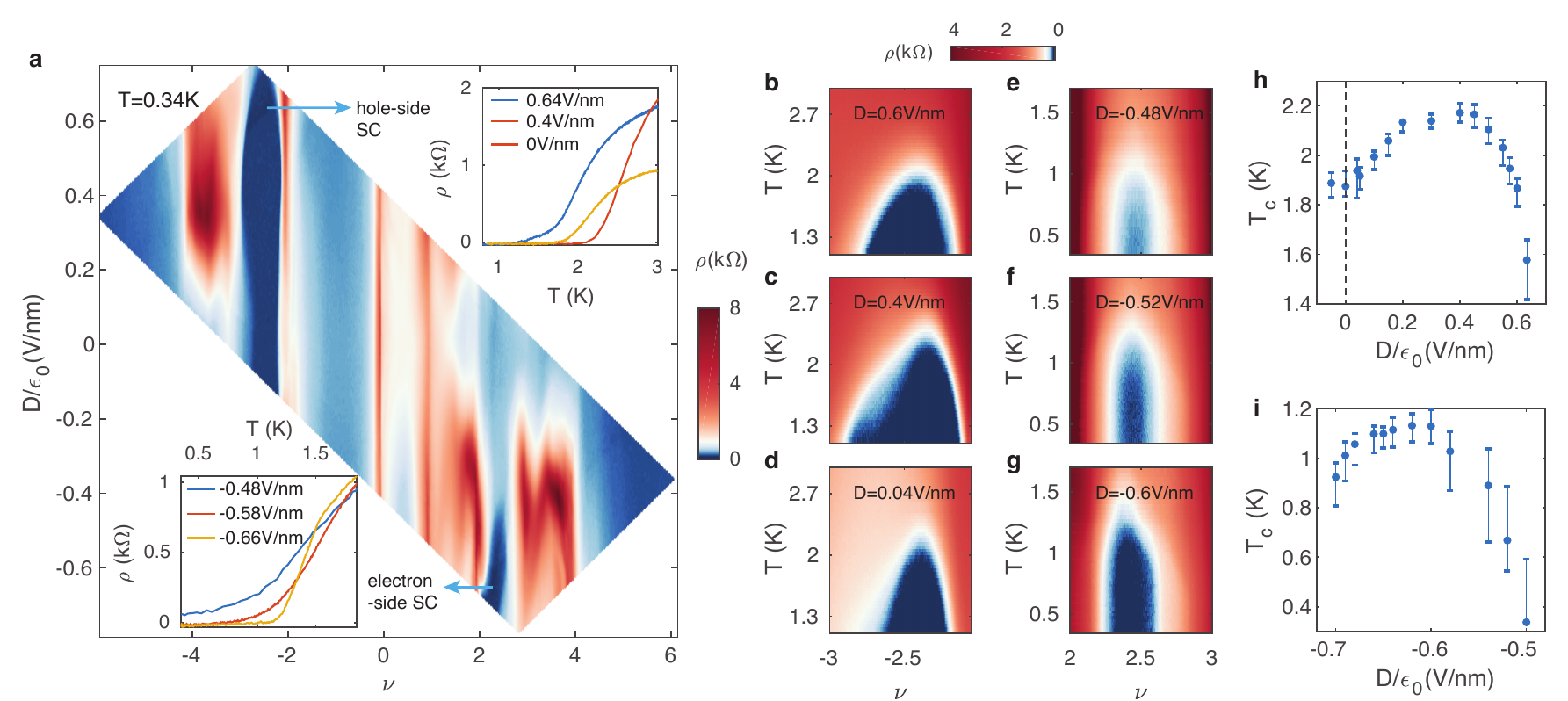}
\caption{\textbf{$\vert$ Electric field tuned superconductivity.} \textbf{a,} $\rho$ map as a function of $\nu$ and $D$. Superconducting regions appear for $\nu<-2$ and $\nu>2$.The upper (lower) inset shows the $\rho$ at the superconducting transition in the hole (electron) region.  \textbf{b-g,} Dome shaped superconducting regions for in the $T-\nu$ plane at different $D$ for $\nu<-2$ (\textbf{b-d}) and $\nu>2$ (\textbf{e-g}). The size and shape of the domes are tuned by the $D$.   \textbf{h,i} $T_c$ as a function of $D$  taken at $\nu=-2.3$ (\textbf{h}) and $\nu=2.45$ (\textbf{i}). $T_c$ is chosen to be the point where $\rho=0.1\rho_N$ and error bars correspond to $0.05\rho_N$ and $0.15\rho_N$.}
\end{figure}

\begin{figure}
\centering
\includegraphics[width=0.5\textwidth]{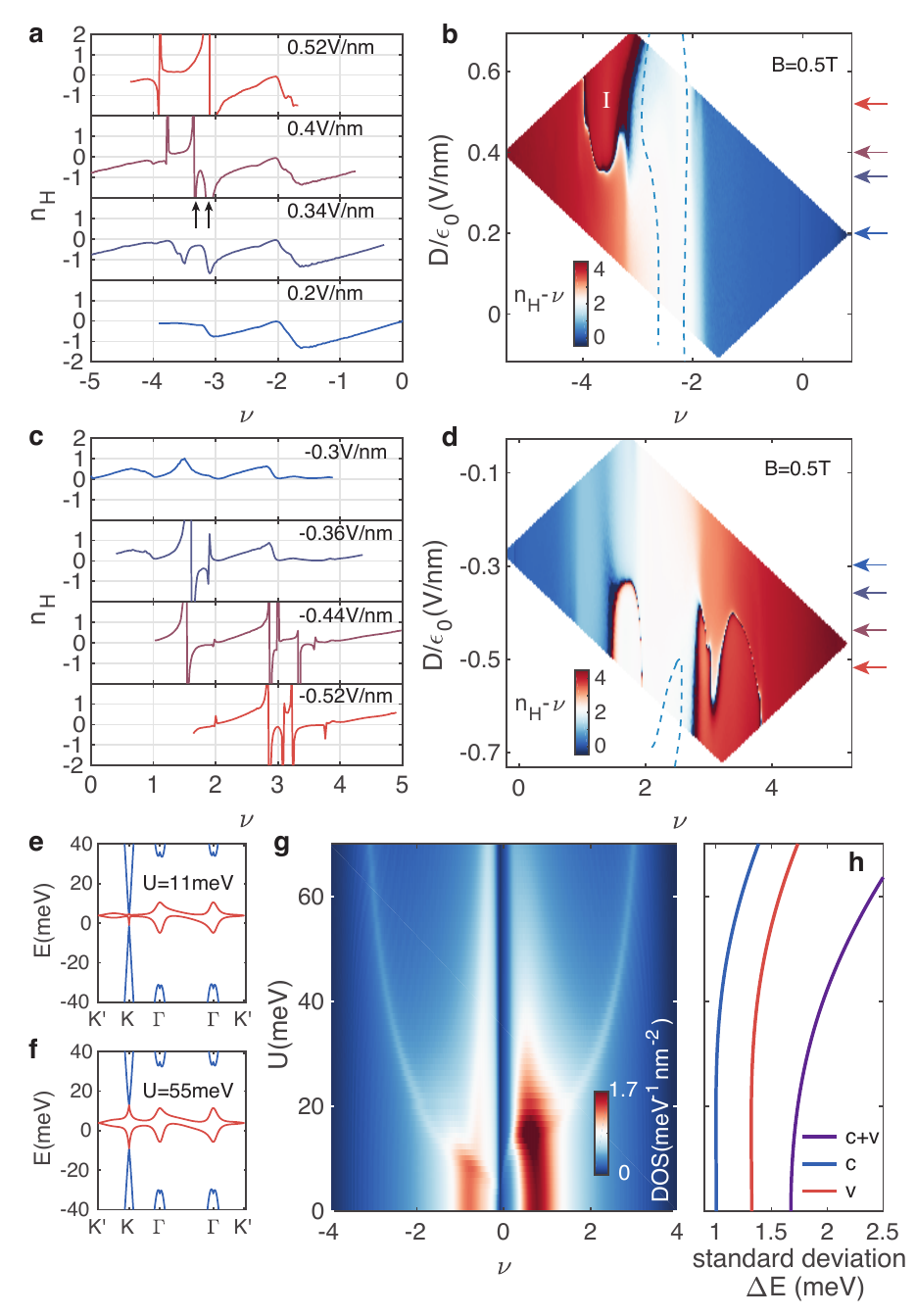}
\caption{\textbf{$\vert$ Hall data and band structure.} \textbf{a,} Hall density, $n_H$ at $0.$5~T at several $D$. Several resets in $n_H$ are visible at flavour symmetry breaking boundaries. Sign reversal vHS with flavor symmetry breaking is marked by vertical arrows  \textbf{b,}Subtracted hall density $n_H-\nu$ as a function of $\nu$ and $D$ in the same region. Showing the symmetry broken regions. Lines mark the locations of the line cuts in \textbf{a}.The $n_H-\nu=4$ region marked by I crowds out the superconducting region at large $D$. \textbf{c}, and \textbf{d}.The same as \textbf{a} and \textbf{b} but near the $\nu=2$ superconducting state. \textbf{e-f,} Theoretical band structures at small (\textbf{e}) and large (\textbf{f}) $U$. The major change in the band structure is a splitting at the $K$ point. \textbf{g,} Calculated density of states as a function of $\nu$ and $U$ with no flavor symmetry breaking. Prominent vHS diverge from the large flat band density of states at large $U$. \textbf{h,} $U$ induced Increase in the flat band width displayed as the standard deviation of the conduction (c), valence (v), and combined (c+v) flat bands as a function of $U$.} 
\end{figure}









\end{document}


\maketitle

\subsection*{S1: Device fabrication and characterization} 
Our twisted trilayer graphene (TTG) device TMB (device names used in Fig.~1) and the twisted bilayer graphene (TBG) device TM have both top and bottom graphite gates. The TBG device MB is controlled by top graphite gate and silicon back gate. The van der Waals heterostructure stack for making the devices consists of 8 layers of two-dimensional materials in the order of hBN, few-layer graphite, hBN, mono-graphene, mono-graphene twisted with angle $\theta$, mono-graphene twisted with angle $-\theta$, hBN and few-layer graphite. The stack was prepared using the dry transfer method, similar to the procedures introduced in most published literature on twisted graphene devices. We make stamps consisting polycarbonate (PC) polymer and polydimethylsiloxane and pick up each layer sequentially. The temperature is kept under 180~$^\circ$C through out the transfer process. We find that generally graphene flakes with large area (e.g., 70~$\mu m$ by 70~$\mu m$) give higher yield in making twisted graphene samples. In order to minimize the movement of graphene flakes during transfer processes, we use an atomic force microscope (Asylum Cypher S) to precut the graphene flakes. For this, we follow the general procedure described in reference \cite{Sai.20c}, using a platinum doped AFM cantilever and contact mode. An 100kHz AC bias of 30V is applied to the cantilever during cutting. We find that this AC bias is critical but its exact role in cutting is currently unknown. The stack is deposited on top of a 300-nm SiO$_2$/Si substrate that has evaporated gold alignment marks on it. Alignment marks are made beforehand so  the twisted sample is not subject to the high temperature of the evaporation process before being etched. Three Hall bar devices were fabricated in the regions of TMB, TM and MB following the standard e-beam lithography and dry etch procedures. 

The transport data was measured at 17.7~Hz using the standard lock-in technique, with a 0.5--1~mV voltage bias and a current-limiting-resistor of 180~k$\Omega$ connected in series with the sample, which limits the current in the sample to an upper bound of 5--10~nA. The sample is connected to the cryostat probe through an RC filter to reduce noise.

We calculate twist angles using two independent methods. The first is to use the geometric capacitance between the twisted samples and gates. The carrier density is determined by the top gate voltage $V_t$ and the bottom gate voltage $V_b$ through $n=c_tV_t+c_bV_b$, where $c_t$ ($c_b$) is the capacitance between the top (bottom) gate and the sample , and it can be directly calculated $c_{t(b)}=\kappa\epsilon_0/d_{t(b)}$. $\kappa$ is the dielectric constant for hBN and is usually taken as 3.9. $\epsilon_0$ is the vacuum permittivity. $d_{t(b)}$ is the top (bottom) hBN thickness. Using the  resistivity, $\rho$, versus gate voltage $V_{t(b)}$ at zero magnetic field, we can associate the resistive peaks with integer fillings of the moir\'e bands, and therefore obtain the gate voltage, or equivalently using the above formulae the carrier density $n_s$ for full filling at $\nu=4$. This carrier density corresponds to 4 electrons per moir\'e unit cell $n_s=4/A_m$, where $A_m$ is the moir\'e unit cell area and is connected to the small twist angle by $A_m=\frac{\sqrt{3}a^2}{2\theta^2}$, where $a$ is the lattice constant for graphene. The main uncertainty in this method comes from the uncertainty in the value of the dielectric constant $\kappa$ and the finite width of the integer-filling resistive features. The second method uses the Landau fan diagrams shown in magnetotransport data. By comparing the longitudinal resistivity data with the Hall conductance, we can assign each line in the Landau fans with a Chern number $C$ so that $\sigma_{xy}=Ce^2/h$, where $e$ is the electron charge and $h$ is the Planck's constant. The slopes of the lines in Landau fans are connected to the Chern numbers through $BA_m/\phi_0=Cn/n_s+s$, where $B$ is magnetic field, $\phi_0=e/h$ is the magnetic flux quantum, $s$ is the filling fraction from which the Landau fan emanates. The main uncertainty in this method comes from how well the slopes are fit. This gives an uncertainty of $\pm0.02^\circ$ in calculating the angle.

\subsection*{S2: Band structure calculation and DOS} 

In this section we discuss the single particle band structure of magic angle twisted trilayer graphene, shown in Figs.~1c and 4e,f.  The density of states was also plotted in Fig.~4g.  The band structure was computed from the trilayer analogue \cite{Kha.19} of the Bistritzer-Macdonald model \cite{Bis.11} of twisted bilayer graphene.  In this case, however, the in-plane displacement between layers matters. As shown in Ref.~\cite{Kha.19}, the Hamiltonian can be brought to a form where only the relative displacement between the top and bottom layers appears.  We denote this distance $\bd$. For a single spin and graphene valley, the Hamiltonian is
\begin{equation}
    H(\bd) = \begin{pmatrix} -i v \bsigma_{\theta/2} \cdot \bnabla & T(\br - \bd/2) & 0 \\
        T^\dag (\br - \bd/2) & -iv \bsigma_{-\theta/2} \cdot \bnabla & T^\dag(\br + \bd/2) \\
    0 & T(\br + \bd/2) & -i v \bsigma_{\theta/2} \cdot \bnabla \end{pmatrix}. 
    \label{ham}
\end{equation}
Here, $\bsigma_{\theta/2} = e^{-\frac{i}{4} \theta \sigma_z} (\sigma_x, \sigma_y ) e^{\frac{i}{4} \theta \sigma_z}$, $v$ is the graphene Fermi velocity, and 
\begin{equation}
    \begin{aligned}
        T(\br) & = \begin{pmatrix} w_0 U_0(\br) & w_1 U_1(\br) \\ w_1 U^*_1(-\br) & w_0 U_0(\br) \end{pmatrix}, \\
        U_0(\br) & = e^{-i \bq_1 \cdot \br } + e^{-i \bq_2 \cdot \br } + e^{-i \bq_3 \cdot \br },  \\
        U_1(\br) & = e^{-i \bq_1 \cdot \br } + e^{i \phi}e^{-i \bq_2 \cdot \br } + e^{-i \phi}e^{-i \bq_3 \cdot \br },
        \label{tunnel}
    \end{aligned}
\end{equation}
with $\phi = 2\pi/3$.  The vectors $\bq_i$ are $\bq_1 = k_\theta(0,-1)$ and $\bq_{2,3} =k_\theta(\pm \sqrt{3}/2,1/2)$.  The wavevector $k_\theta = 2k_D\sin \frac{\theta}{2}$ is the moir\'{e} version of the Dirac wavevector $k_D= 4\pi/3a_0$, where $a_0$ is the graphene lattice constant. For the other graphene valley, the Hamiltonian is the complex conjugate of \eqref{ham}.  
The spectrum of $H(\bd)$ depends strongly on $\bd$.  However, Ref. \cite{Car.20} finds that $\bd = 0$ has the lowest energy due to relaxation effects, and that the system is likely to slide into this configuration naturally.  We therefore focus on $\bd = 0$ which corresponds to AA stacking between the top and bottom layers.

For $\bd = 0$, the Hamiltonian has a symmetry under exchanging the top and bottom layer.
\begin{equation}
    M_z = \begin{pmatrix} 0 & 0 & 1 \\ 0 & 1 & 0 \\ 1 & 0 & 0 \end{pmatrix}.
\end{equation}
We may then consider separately the Hamiltonian in the $M_z = \pm1$ sectors.  For $M_z = +1$ we find a TBG Hamiltonian 
\begin{equation}
    H_+ = \begin{pmatrix} -i v \bsigma_{\theta/2} \cdot \bnabla & \sqrt{2} T(\br) \\
    \sqrt{2} T^\dag(\br) & -iv \bsigma_{-\theta/2} \cdot \bnabla \end{pmatrix},
    \label{tbgsector}
\end{equation}
where the tunneling is $\sqrt{2}$ times stronger.  On the other hand for $M_z = -1$ we obtain ordinary graphene
\begin{equation}
    H_- = -iv \bsigma_{+\theta/2} \cdot \bnabla.
    \label{graphenesector}
\end{equation}
Here, the ordinary graphene electrons come from the top and bottom layers only and the Dirac cone is centered around the moir\'{e} $K$ point.  Similarly in the other graphene valley the Dirac cone is centered at the moir\'{e} $K'$ point. Thus, for this system we expect that when the angle is $\sqrt{2}$ times the TBG magic angle, we will obtain flat bands from \eqref{tbgsector} together with a Dirac cone from \eqref{graphenesector}.  This band structure is depicted in Fig.~1c. with parameters $\theta = 1.55^\circ$, $w_1 = 110 \rm{meV}$, and $\kappa = w_0/w_1$.

A nonzero displacement field  mixes the TBG and graphene sectors by breaking $M_z$; its effect is largest at the K point where the bands intersect. There, the two Dirac points near charge neutrality, one from each of the graphene and TBG subsystems, split and hybridize so that there is one above zero energy and one below zero energy.  These Dirac points are still protected by inversion combined with time reversal which acts as $H(\br) \to \sigma_x H^*(-\br) \sigma_x$ and is a symmetry when $\bd = 0$.  Band structures with nonzero displacement fields are shown in Fig.~4e,f. with the same parameters as Fig.~1c.

\begin{figure}
    \centering
    \includegraphics[width = 0.9\textwidth]{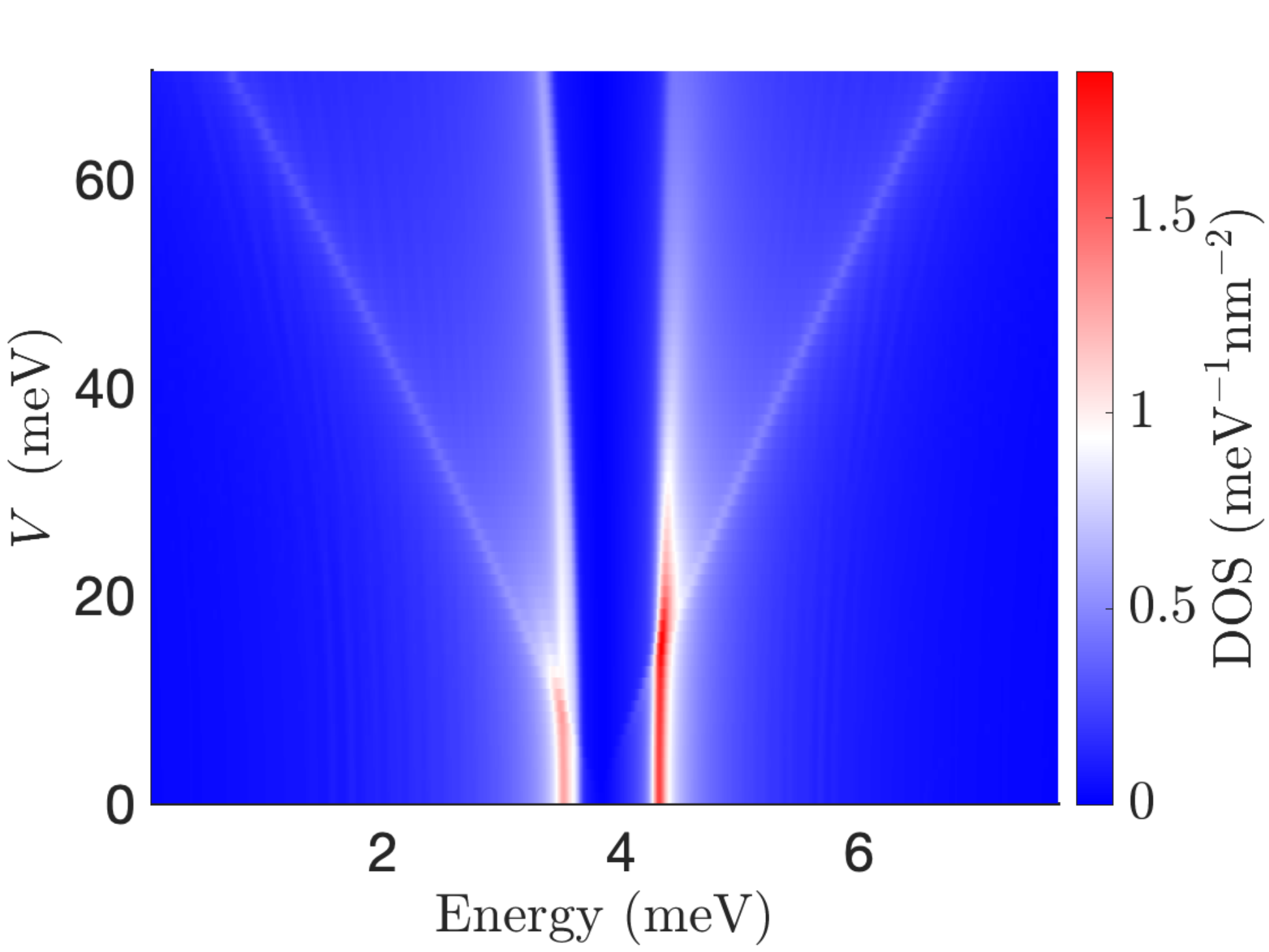}
    \caption{ \textbf{Density of states versus energy and displacement field.} Similar to Figure 4g in the main text, one sees two flat bands that spread out after a sufficiently large displacement field is applied.  Prominent van Hove singularities are visible in white and spread out with increasing displacement field.}
    \label{fig:dosenergy}
\end{figure}

The density of states is shown in Fig.~4g in the main text. It is obtained from the band structure of the Hamiltonian \eqref{ham} by a gaussian-smoothing over energy levels with standard deviation $0.03$ meV. Here we also include the density of states plotted versus energy instead of filling, see Fig.~S\ref{fig:dosenergy}.

\subsection*{S3: Sample homogeneity} 
Fig.~S2 shows a comparison of $\rho$ versus n measured with $V_{\text{BG}}=0$ for different pair of contacts. From Fig.~S2A to E, the blue circles in the device image illustrate the pair of contacts, labeled P1 - P5, used for measuring the data on the right. Red dashed lines label the resistive states at integer fillings $\nu=$-2, 0, 1, 2, 3. It can be seen that for different contacts, the red dashed lines are slightly misaligned with respect to each other, indicating that the regions between the contacts have different angles. We calculated the angle for P2 using quantum oscillation. Then using their relative ratio of full filling densities, we obtained angle for each pair of contacts. The  measured angles are $\theta_1=1.552^\circ$,$\theta_2=1.567^\circ$, $\theta_3=1.567^\circ$, $\theta_4=1.572^\circ$, and $\theta_5=1.572^\circ$ all with uncertainty $\pm0.02^\circ$. Although the third digit in the angles maybe seem meaningless given the magnitude of the uncertainty, they indicate the relative angle difference between different pairs of contacts, which has smaller uncertainty. There are  double peak features in P5 indicating a region of $\theta=1.61^\circ$. None of the presented data was taken in this more diordered region. We note that over the majority of the sample, the angle is extremely uniform changing less that $0.2^\circ$. The angle gradually becomes larger from the left to the right. And there is more angle disorder on the right side of the sample. The superconductivity is strongest at the left end of the sample with $\theta_1=1.552^\circ$, and the majority of the presented data was taken in this region.

\begin{figure}
\centering
\includegraphics[width=0.9\textwidth]{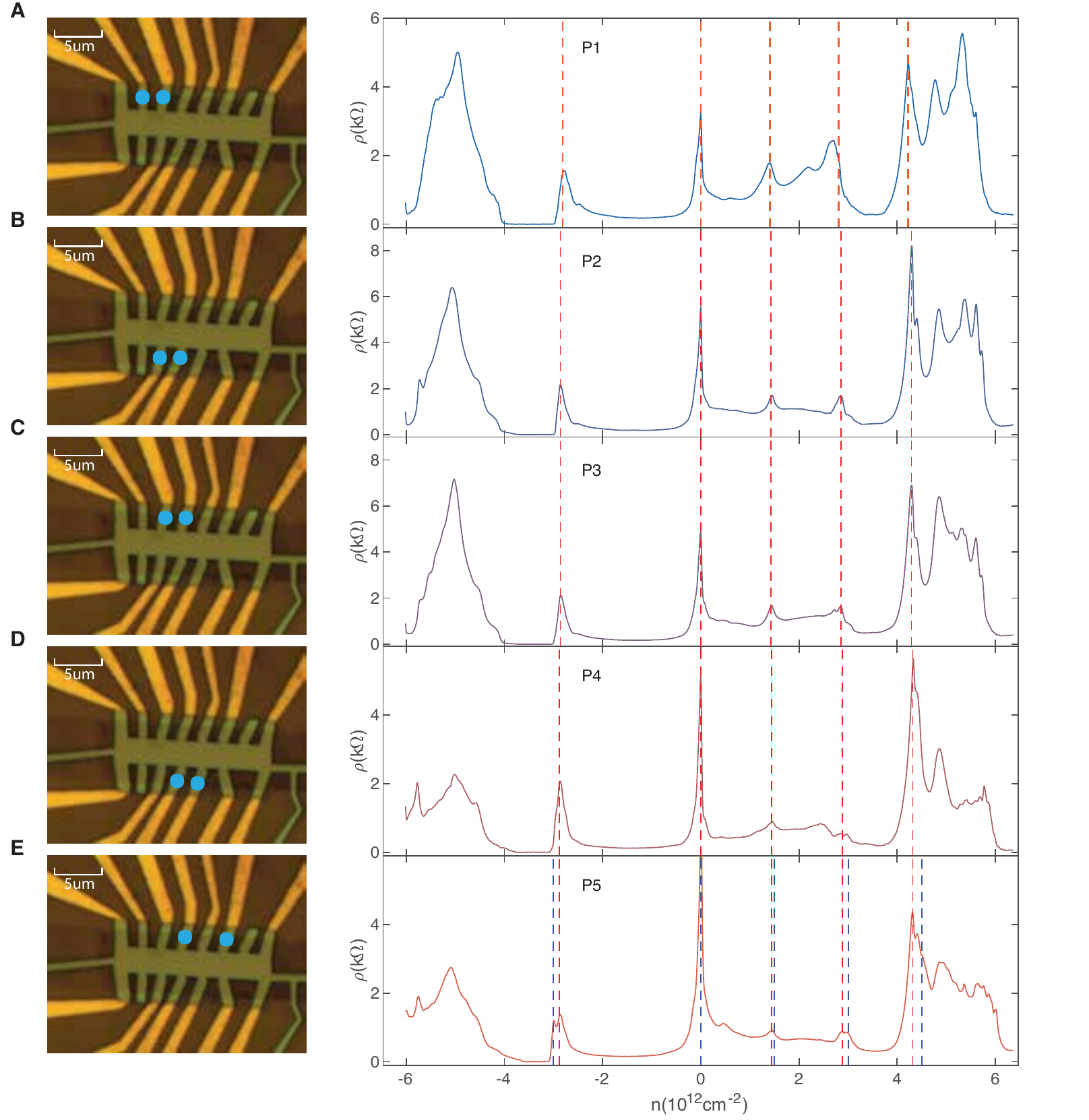}
\caption{\textbf{Angle inhomogeneity.} The blue circles mark which pair of contacts are measured for the data in each figure. Red dashed lines label integer fillings from left to right $\nu=$-2, 0, 1, 2 and 3.}
\end{figure}

\subsection*{S4: Fan diagram comparison} 
Fig.~S3A shows inverse Hall resistivity $1/\rho_{xy}$ as a function of $\nu$ and B at $V_{\text{BG}}=0$ and T=340~mK. This is complementary to the $\rho(B,n)$ data shown in Fig.~1f. To better illustrate the quantization values, in Fig.~S3B, we overlay the $1/\rho_{xy}(B,n)$ data with contours that have values $(l\pm0.5)e^2/h$, where $l$ is an integer between 0 and 15. We can see large area of $1/\rho_{xy}=-2e^2/h$ near $\nu=-4$ and $2e^2/h$ near $\nu=4$. From the single particle band calculation we know that when a displacement field is applied, the originally independent Dirac cone and flat bands at zero displacement field  mix resulting in two Dirac cones splitting to higher and lower energy respectively. The large regions of quantized reverse Hall resistivity are likely from the quantum Hall states of these Dirac cones. Fig.~S2C is a schematic of the quantum Hall structure observed in Fig.~1f. Emanating from the charge neutrality, the main sequences are $C=-2, -6, -10,\dots$ on the hole doped side and $C=2, 6, 10$ on the electron doped side. At higher magnetic field, between $C=-2$ and $C=-6$, symmetry breaking states with $C=-3, -4$ and $-5$ emerge and the sequence $C=-14,-18,-20$ transitions into $C=-12, -16, -20$.

Fig.~S4 shows the fan diagram at zero displacement field. The Landau fan sequences emerging from neutrality, $\nu=\pm2$ are similar to that observed in the $V_{BG}=0$ fan.  Interestingly, in low field range, we observe ``arc-like" features or a coexisting quantum oscillation structure distinct from those of the flat bands. We argue these are in fact the quantum Hall states of the additional Dirac cone. Fig.~S5A shows $\rho$ and Hall conductance $\sigma_{xy}$ as a function of inverse magnetic field at $\nu=-3.82$, where the arcs are prominent. We see clear quantum oscillations, displayed as the equally distanced minima in $\rho$. The $\sigma_{xy}$ values corresponding to the first two minima are quantized to $-2e^2/h$ and $-6e^2/h$, the same as the Dirac Landau level sequence. $\sigma_{xy}$ for the other  minima are not well quantized. Most likely  because they appear at lower magnetic fields and are not well developed. Fig.~S5B shows $\frac{d\rho}{dB}$ to make these quantum oscillations more prominent.

The quantum Hall states of the Dirac cone appear as arcs instead of the normal straight lines emanating from $\nu=0$ because the flat bands and the Dirac cones are filled simultaneously. Kinks in the Dirac cone states correspond to changes in the flat band chemical potential  due to strong-interaction induced symmetry breaking as has been observed in TBG\cite{Zon.20,Won.20,park2020flavour}. To confirm that these structures are the Landau fan of the additional Dirac cone, in Fig.~S5C and D, we trace out $\rho$ minima for the visible states which we  assume are the $C=6$, 10 and 14 states for Dirac cone, labeled by the triangle symbols with different colors. With these traces, we obtain the positions in magnetic field $B_6$, $B_{10}$, and $B_{14}$ as a function of $\nu$ for the $C= 6$, 10 and 14 states respectively. According to Diophantine equation $\nu-s=C\phi/\phi_0$ (where s is the filling where the Landau fan emerges from and $\phi=BA$ is the magnetic flux through the sample and $A$ is the sample area), if these structures are quantum Hall states of $C=6$, 10, 14, we expect that $6B_6=10B_{10}=14B_{14}$ even with charge carriers split between the Dirac cone and the flat bands. Fig~S5B and C shows the normalized ratio between $6B_6$, $10B_{10}$ and $14B_{14}$ and indeed they are roughly one, confirming that the states originate from the Dirac cone. We also note that these arcs are not present when the fan diagram is measured with a finite $D$ as shown in Fig. 1 of the main text. This is consistent with the theoretical prediction that a gap opens at the Dirac cone when a displacement field is applied, causing the Dirac cones to not fill until after the flat bands.

\begin{figure}
\centering
\includegraphics[width=0.9\textwidth]{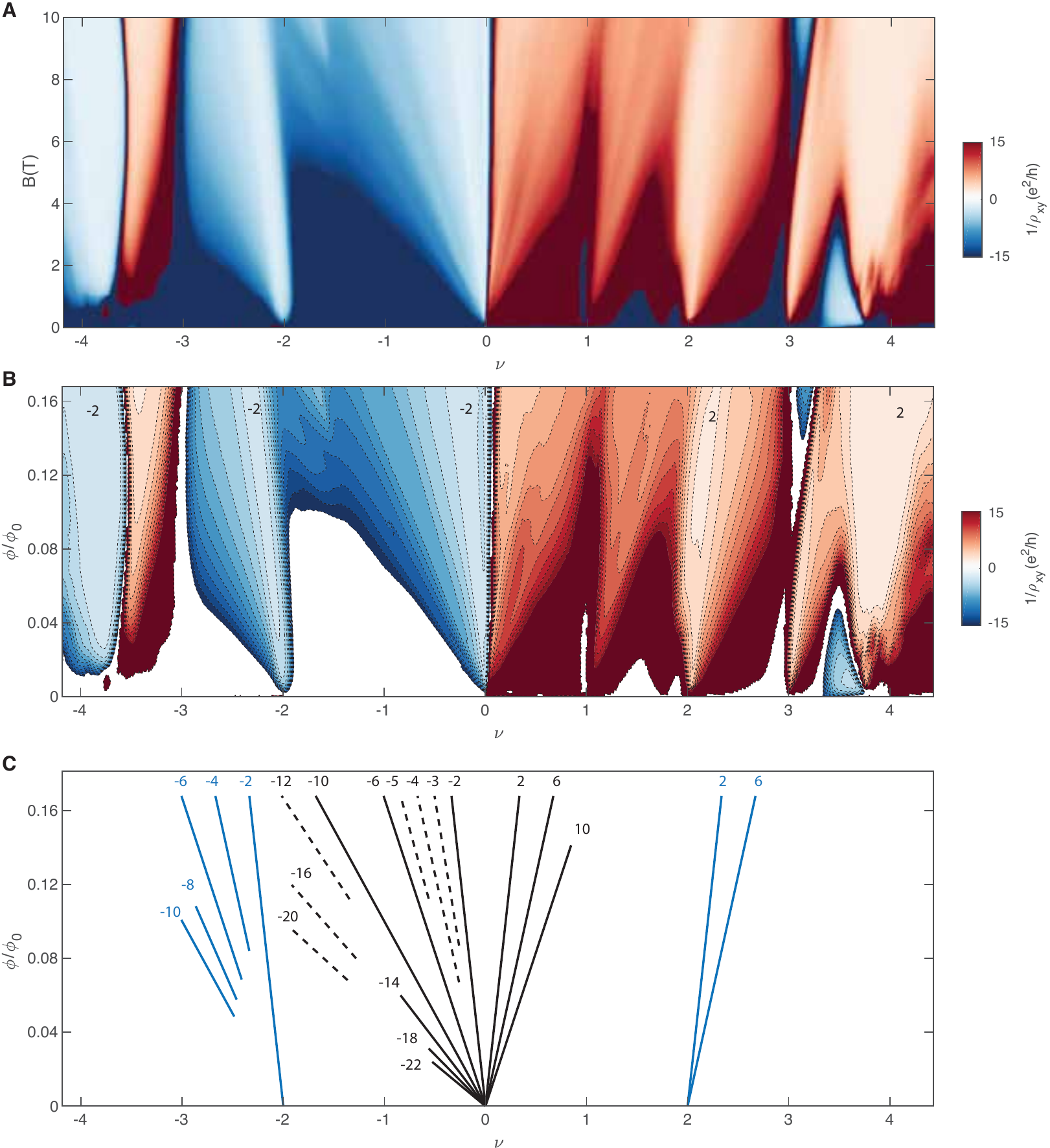}
\caption{\textbf{TTG Hall resistivity at $V_{\text{BG}}=0$ and T=340mK}. \textbf{(A)} $1/\rho_{xy}$ as a function of $\nu$ and B. \textbf{(B)} Contour plot of $1/\rho_{xy}$. Contour values are taken as $(l\pm0.5)e^2/h$, where $l$ is an integer between 0 and 15. Large regions of $1/\rho_{xy}=\pm2e^2/h$ are marked. \textbf{(C)} Schematic of the quantum Hall structure in the Landau Fan diagram at $V_{\text{BG}}=0$.}
\end{figure}

\begin{figure}
\centering
\includegraphics[width=0.9\textwidth]{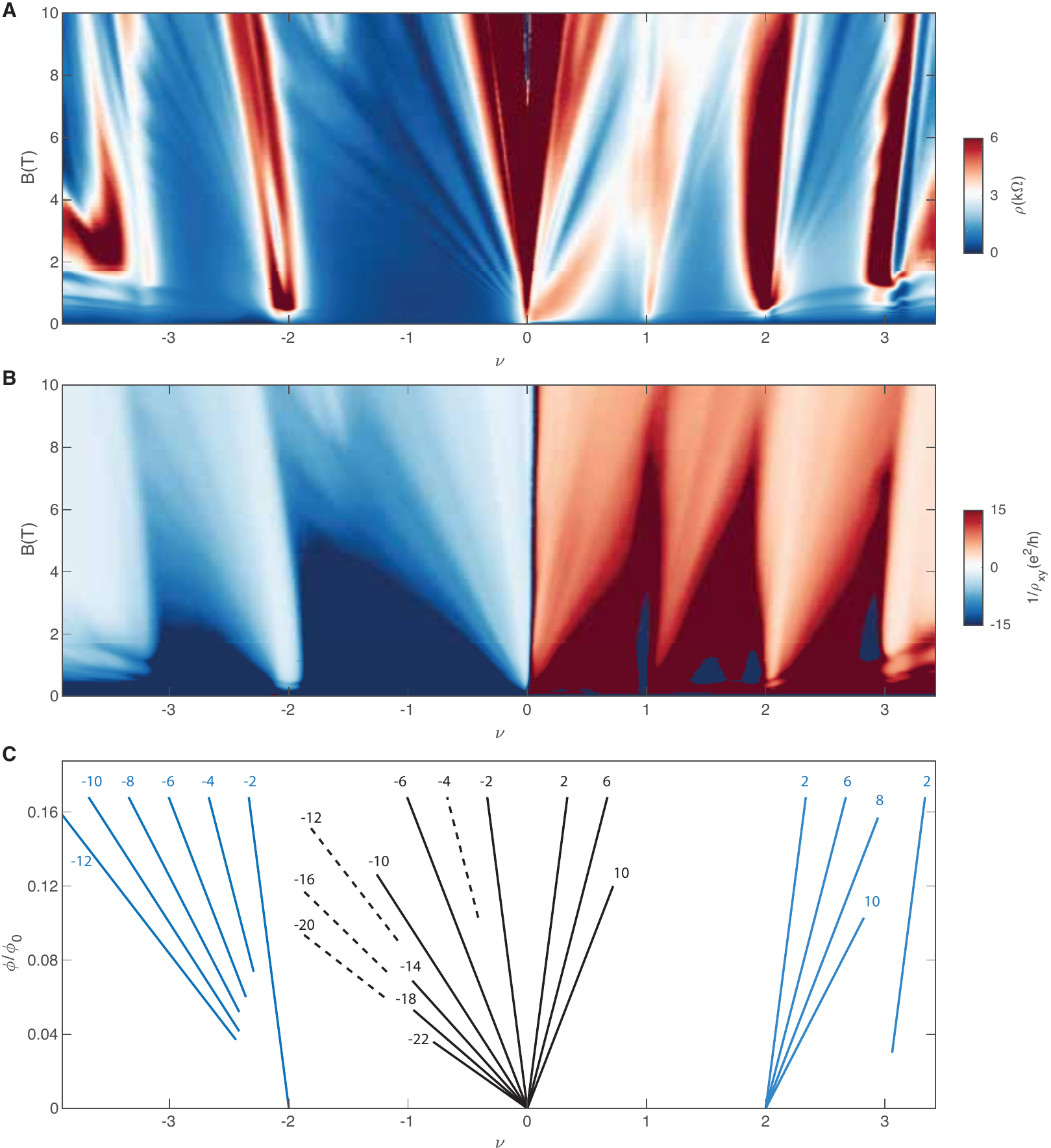}
\caption{\textbf{TTG Landau fan diagrams at $D=0$ and T=340mK.} \textbf{(A)} and \textbf{(B)} $\rho$ and $1/\rho_{xy}$ as a function $\nu$ and B. \textbf{(C)} Schematic of the quantum Hall structure.}
\end{figure}

\begin{figure}
    \centering
    \includegraphics[width=0.9\textwidth]{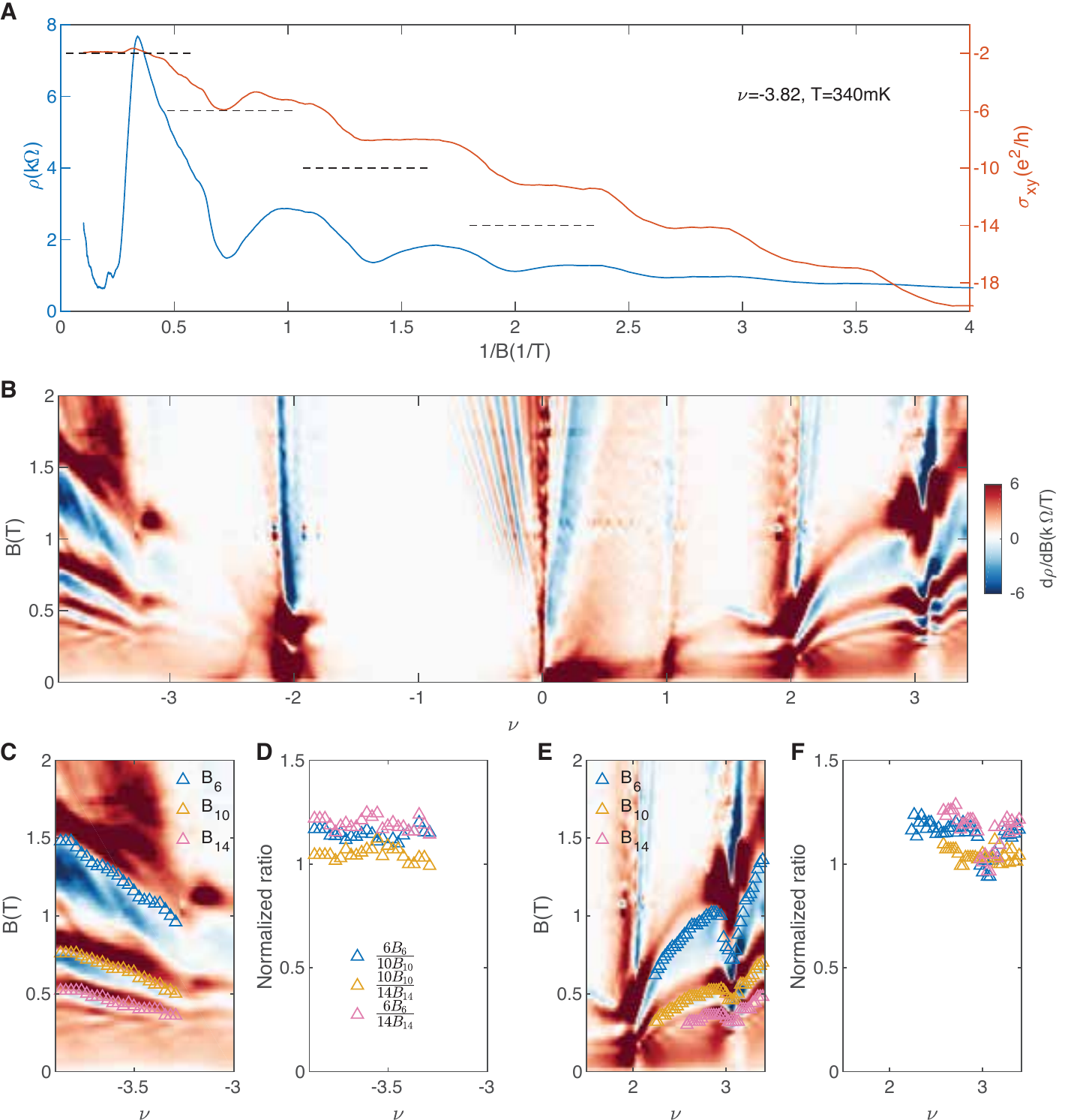}
    \caption{\textbf{Evidence of the additional Dirac cone.}\textbf{(A)}Quantum oscillations at $\nu=-3.82$. Blue curve is longitudinal resistivity $\rho$ and orange curve is Hall conductance $\sigma_{xy}$. Black dashed lines mark quantized conductance values corresponding to sequence -2, -6, -10...\textbf{(B)} $\frac{d\rho}{dB}$ of the $\rho$ data in Fig.~S4A below 2T.\textbf{(C)} and \textbf{(E)} zoom-ins of the oscillations on the hole side and the electron side. Triangles trace $\rho$ minima. $B_{6}$, $B_{10}$ and $B_{14}$ correspond to the Dirac cone Landau fan sequence 6, 10 and 14. \textbf{(D)} and \textbf{(F)} Normalized ratio between the magnetic field values multiplied by the sequence number of different traces. The ratios are all approximately  1. } 
    \label{dirac_evidence}
\end{figure}

\subsection*{S5: Critical temperature and GL coherence length} 
\begin{figure}
\centering
\includegraphics[width=0.6\textwidth]{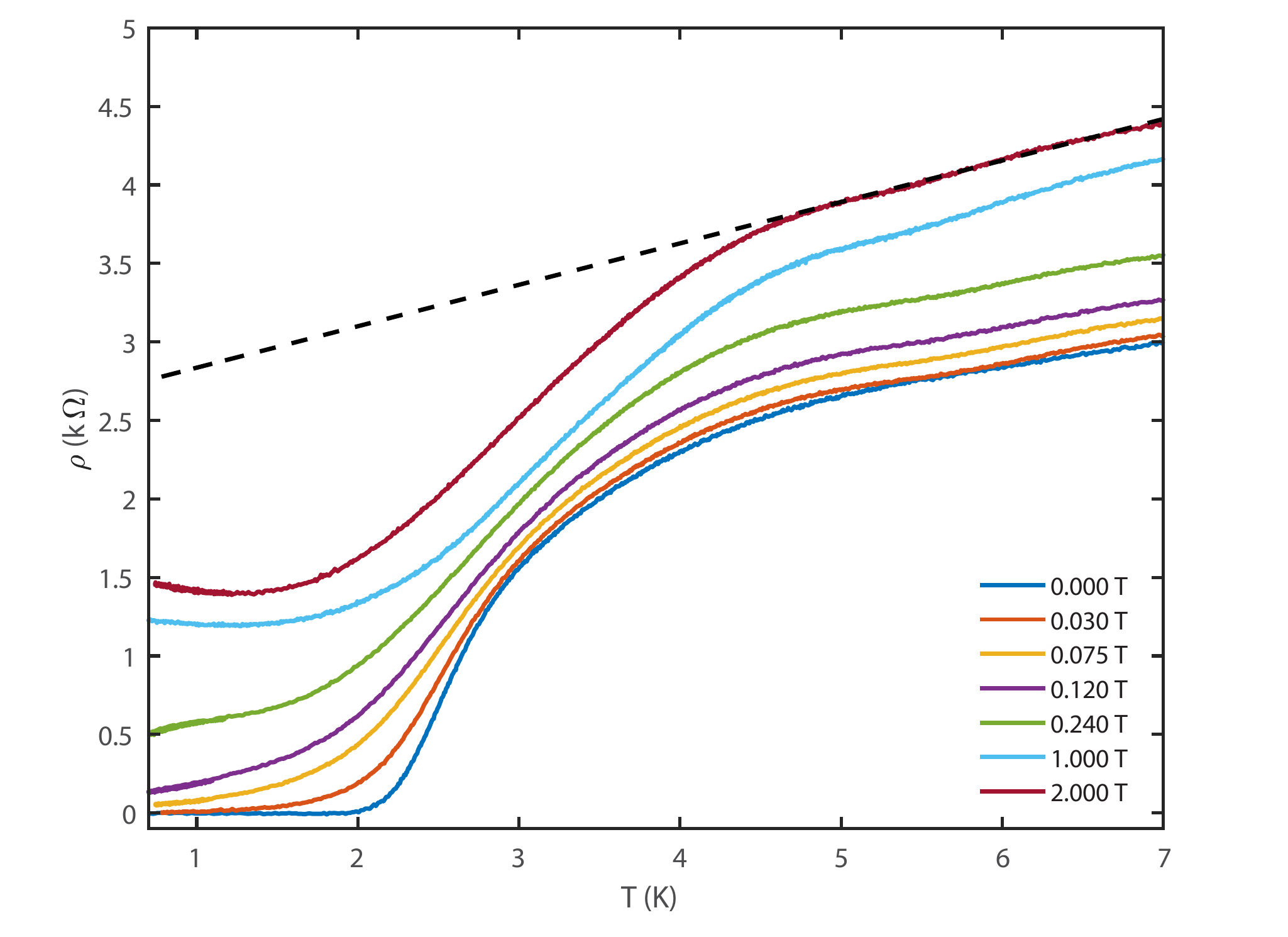}
\caption{\textbf{$\mathbf{T_c}$ Extraction} $\rho(T)$ measured at several different fields at $\nu=-2.3$ and $D/\epsilon_0=0.3$~V/nm. The dashed line is a fit to the normal state resistivity at  2~T.}
\label{fig.Tcextract}
\end{figure}

\begin{figure}
\centering
\includegraphics[width=0.9\textwidth]{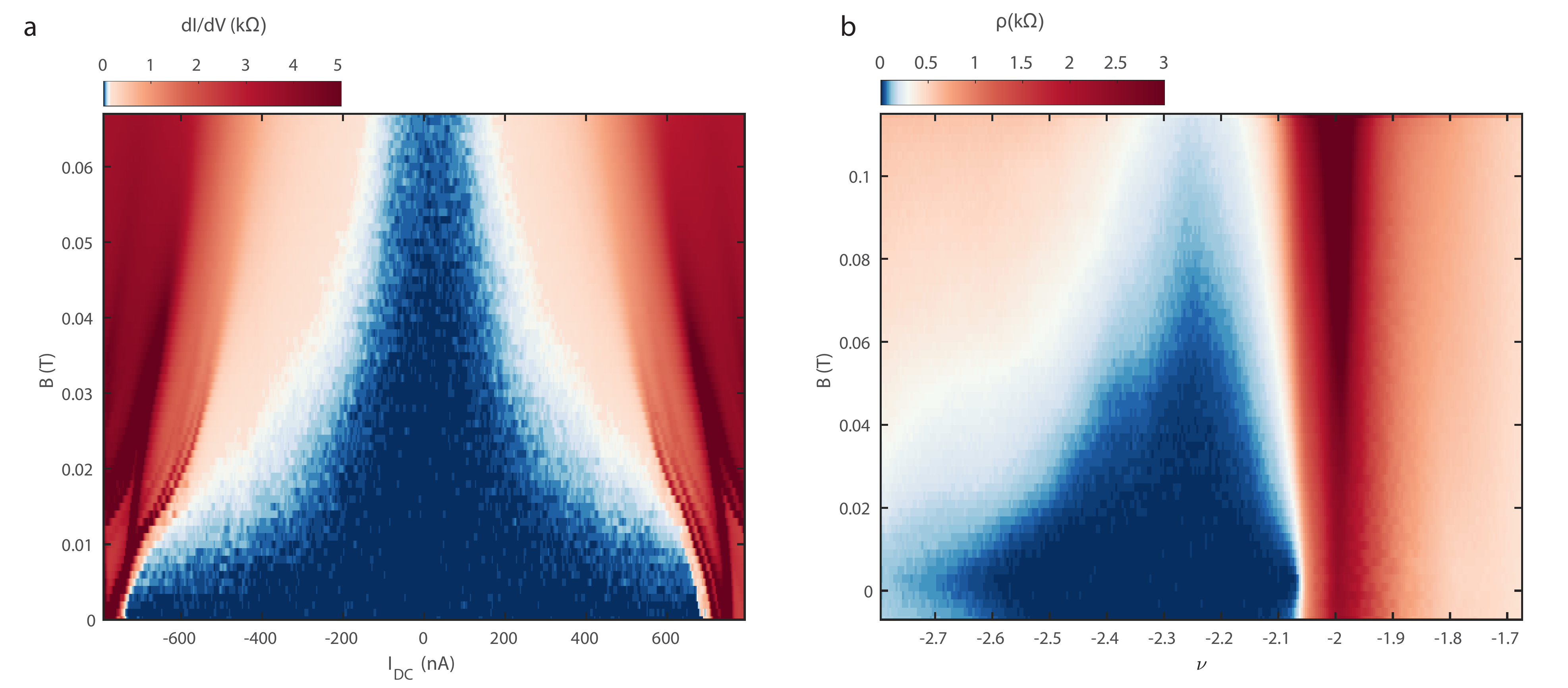}
\caption{\textbf{B dependence} \textbf{a} Differential resistance as a function of DC bias current showing a long low resistance tail extending to large $B$. \textbf{b} Resistivity as a function of $nu$ and $B$ also with low resistivity at larger $B$. }
\label{fig.Bfigs}
\end{figure}

We extract $T_c$ from $\rho(T)$ by extrapolating the normal state resisitivty $\rho_N$ to low temperature by fitting a line to the high temperature linear $\rho$ in the normal state and finding the temperature where $\rho(T)=x\rho_N(T)$, where $x$ is a percentage. An example of this linear fit for data at 2~T is shown in Fig~S\ref{fig.Tcextract} as well as $\rho(T)$ taken at several different magnetic fields. we note that a dip in $\rho$ at low temperature is evident in the data even at high $B$ although $\rho$ does not go to zero. A similar tail of low $\rho$ extending to high field is evident in the $dV/dI$ data and  $\rho(\nu, B)$, shown in Fig.~S\ref{fig.Bfigs}. It is possible that this dip is due to a vortex phase with non-zero $\rho$. 

\begin{figure}
\centering
\includegraphics[width=0.9\textwidth]{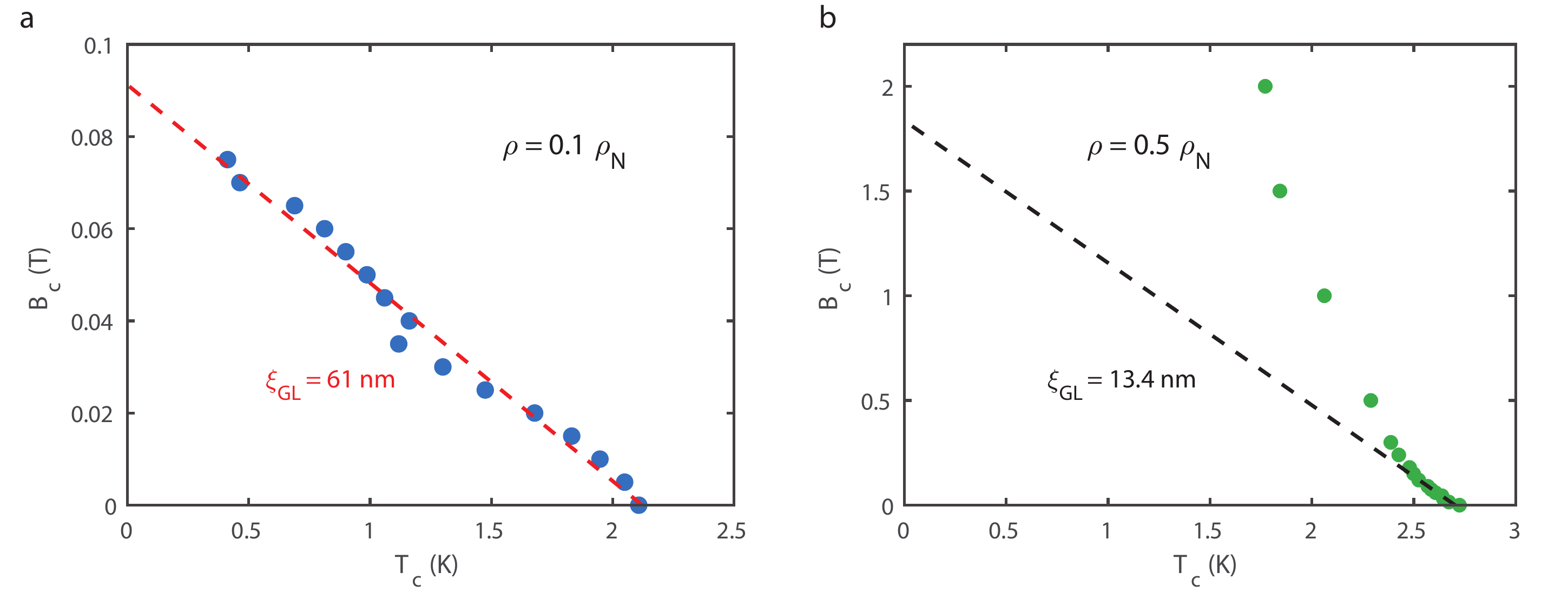}
\caption{\textbf{Coherence Length} $T_c$ determined using $x=0.1$ \textbf{a} and $x=0.5$ \textbf{b}. Dashed lines are fits to GL theory with different $\xi_{\text{GL}}$.}
\label{fig.coh}
\end{figure}

This dip in resistivity results in very different results for $T_c$ depending on the choice of $x$. Fig.~S\ref{fig.coh} shows this difference for $x=0.1$ and $x=0.5$. For $x=0.1$ we find a linear relationship as described by the Ginzburg-Landau (GL)  theory for a two-dimensional superconductor: $B_c=[\Phi_0/(2\pi\xi^2_{\text{GL}})](1-T/T_c)$ where $\xi_{\text{GL}}$ is the GL coherence length \cite{Tin.04}. For $x=0.1$, $\xi_{\text{GL}}=61$~nm. For $x=0.5$ the resulting $T_c$ is much higher and remains above 1.5~K to fields larger than 2~T. It is also very non-linear, although fitting the low field portion gives $\xi_{\text{GL}}=13.4$~nm. We have chosen $x=0.1$ as the standard for this paper as it more clearly defines the region where we observe $\rho=0$ at low temperature, and agrees well with the measured BKT transtion temperature.

\subsection*{S6: Strong-coupling superconductivity} 
\begin{figure}
\centering
\includegraphics[width=1\textwidth]{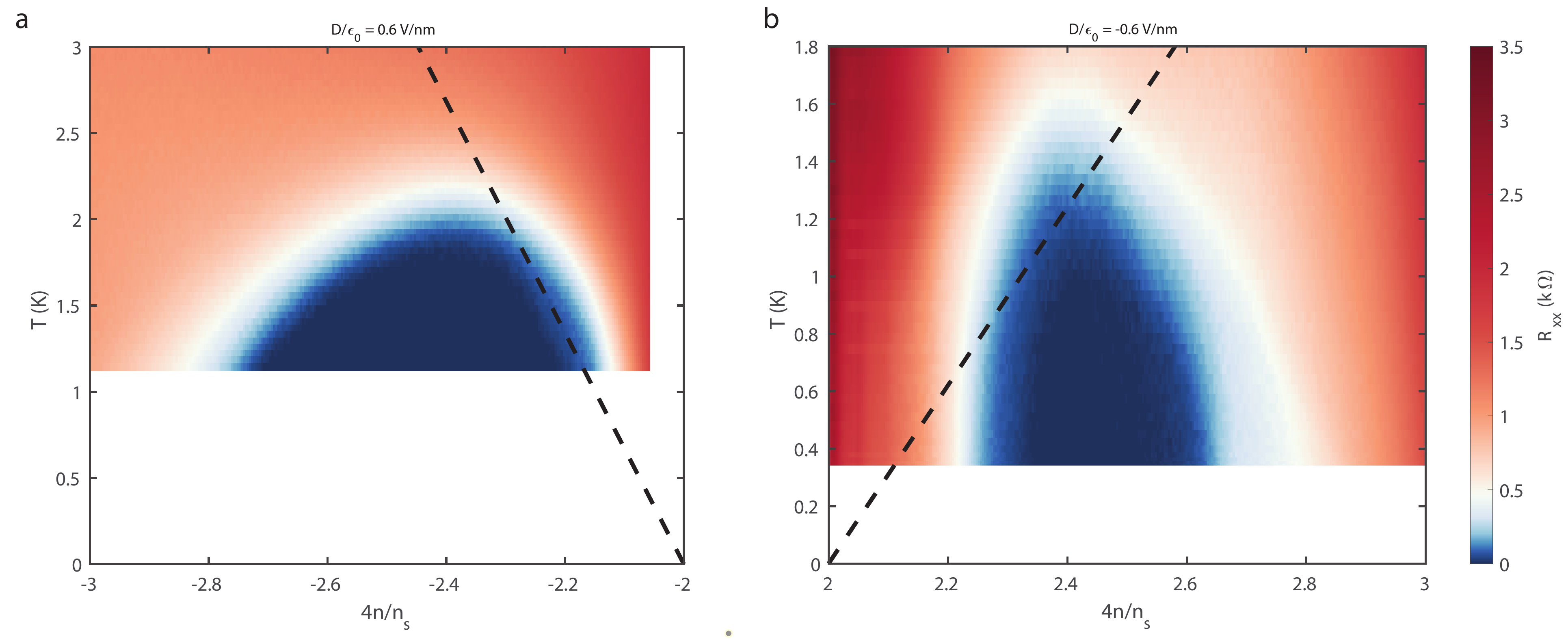}
\caption{\textbf{Strong coupling fit} Examples of typical fits to extract the slope of $T_c$ for comparison with a strong-coupling model of superconductivity. The black dashed lines show fits to $T_c$ near the optimal point on the superconducting domes constrained to pass through $\nu=\pm2$. }
\label{fig.FitEx}
\end{figure}
\begin{figure}
\centering
\includegraphics[width=1\textwidth]{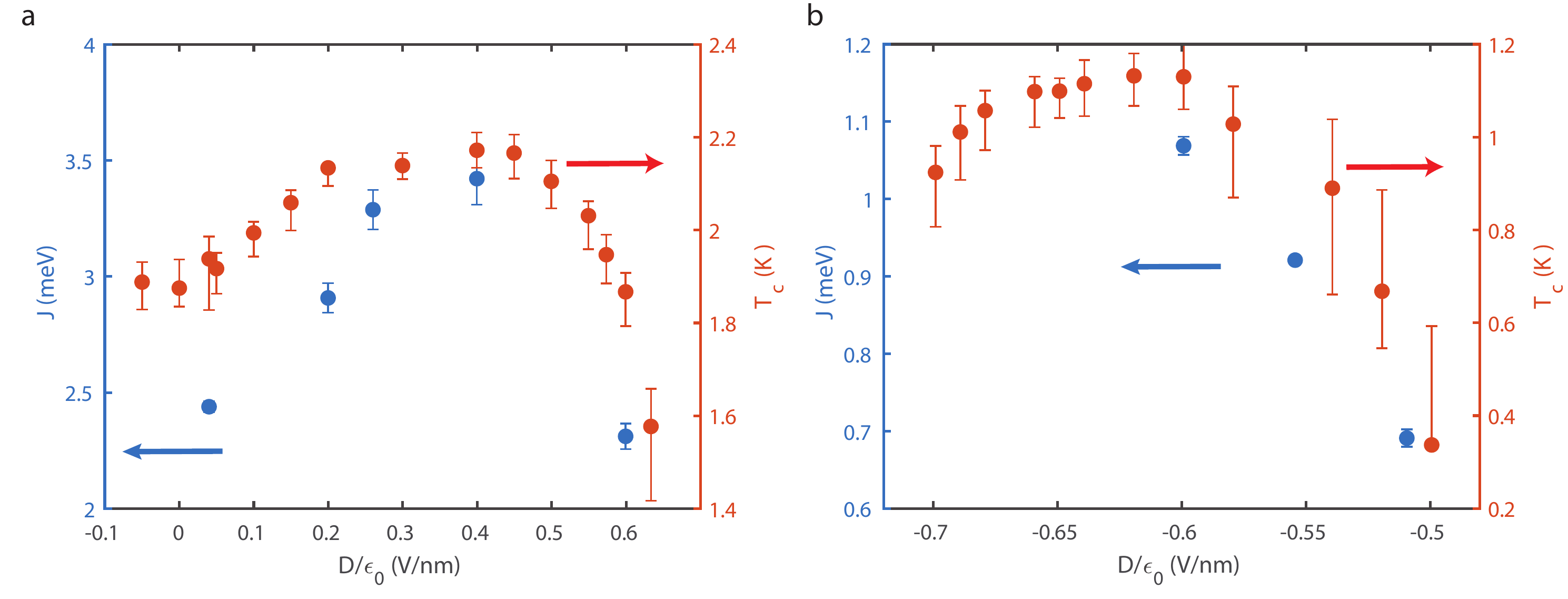}
\caption{\textbf{Extracted pairing scale} Extracted values of $J$ from the fits to the strong-coupling model in the hole (\textbf{a}) and electron (\textbf{b}) regions of superconductivity compared with maximum $T_c$ at each $D$ from Fig. 3 of the main text. We see $J$ and the $T_c$ are correlated.}
\label{fig.J}
\end{figure}

The rapid increase in $T_c$ with doping in addition to the suppression of superconductivity due to the van Hove singularity point towards a strong coupling (BEC) scenario for superconductivity where preformed bosonic charge $2e$ objects are condensed. One such model is the skyrmion model of superconductivity proposed in Ref.~\cite{Kha.20} where such bosonic charge $2e$ objects were proposed to be topological skyrmion textures in some pseudospin variable. Regardless of the actual mechanism, a strong-coupling BEC superconductor obtained by condensing charge $2e$ objects whose density is $\nu_{2e}$ and mass is $M_{2e}$ is characterized by the critical temperature  \cite{NelsonKosterlitz}
\begin{equation}
    k_B T_c = \frac{\nu_{2e} \pi\hbar^2}{2A_MM_{2e}}=\frac{\nu_{2e} J}{2},
\end{equation}
Here, we take the filling fraction of the charge $2e$ objects $\nu_{2e}$ to be equal to half the filling fraction measured relative to half-filling $\nu = \pm 2$. $A_M$ denotes the area of the moir\'e unit cell and $J$  is an effective pairing scale. We can use this formula to extract $J$ from our data by fitting $T_c(\nu)$ near its maximum, where we expect this formula to apply. Fig.~S\ref{fig.FitEx} Shows examples of these fits in the electron and hole superconducting domes superimposed onto the dome resistivty. Since the superconductivity appears only in the flavour symmetry broken regions where $n_H-\nu=2$, where the relevant carrier filling fraction is related to $\nu=\pm2$, we constrain the fits so that $T_c(\nu=\pm2)=0$. The resulting fits agree reasonably well with our data and correspond to values of $J$ between 2.5 and 3.5~meV on the hole side and 0.6 and 1.2~meV on the electron side. This is roughly of the same order as the  coupling scale predicted theoretically \cite{Kha.20}. Moreover, as shown in Fig.~S\ref{fig.J}, we find that $J$ is correlated with the maximum $T_c$ at a given $D$ as we would expect if $J$ is a measure of the pairing strength. We note that based on our extracted values of $J$ we calculate $M_{2e}\sim m_e$ for the hole superconductivity and $M_{2e}\sim 3 m_e - 5 m_e$ for the electron superconductivity.


\pagebreak
\bibliography{SI_TTG.bib}
\bibliographystyle{naturemag_noURL}